\documentclass[twocolumn,floatfix,superscriptaddress,citeautoscript,10pt,aps,pra]{revtex4-2}
\usepackage{graphicx}
\usepackage{xspace}
\usepackage{color}
\usepackage{siunitx}
\usepackage{xcolor}
\usepackage{blindtext}
\usepackage{acronym}
\newcommand{\siv}{\ac{siv}\xspace}

\newcommand{\tief}[1]{_\mathrm{#1}}

\newcommand{\first}{1$^\mathrm{st}$~}
\newcommand{\second}{2$^\mathrm{nd}$~}
\newcommand{\third}{3$^\mathrm{rd}$~}
\newcommand{\nd}{\ac{nd}\xspace}
\newcommand{\opo}{\ac{opo}\xspace}

\DeclareSIUnit{\counts}{c}
\expandafter\def\expandafter\normalsize\expandafter{%
    \normalsize
  \setlength\abovedisplayskip{4pt}
  \setlength\belowdisplayskip{4pt}
  \setlength\abovedisplayshortskip{4pt}
  \setlength\belowdisplayshortskip{4pt}
}

\begin{document}

\title{Hybrid quantum photonics based on artificial atoms placed inside one hole of a photonic crystal cavity}%
\author{Konstantin G. Fehler}
\thanks{Authors contributed equally to this work.}
\affiliation{Institut f\"ur Quantenoptik, Universit\"at Ulm, D-89081 Ulm, Germany}
\affiliation{Center for Integrated Quantum Science and Technology (IQst), Ulm University, Albert-Einstein-Allee 11, D-89081 Ulm, Germany}
\author{Lukas Antoniuk}
\thanks{Authors contributed equally to this work.}
\affiliation{Institut f\"ur Quantenoptik, Universit\"at Ulm, D-89081 Ulm, Germany}
\author{Niklas Lettner}
\thanks{Authors contributed equally to this work.}
\affiliation{Institut f\"ur Quantenoptik, Universit\"at Ulm, D-89081 Ulm, Germany}
\author{Anna P. Ovvyan}
\affiliation{Institute of Physics and Center for Nanotechnology, University of Münster, D-48149 Münster, Germany}
\author{Richard Waltrich}
\affiliation{Institut f\"ur Quantenoptik, Universit\"at Ulm, D-89081 Ulm, Germany}
\author{Nico Gruhler}
\affiliation{Institute of Physics and Center for Nanotechnology, University of Münster, D-48149 Münster, Germany}
\author{Valery A. Davydov}
\affiliation{L.F. Vereshchagin Institute for High Pressure Physics, Russian Academy of Sciences, Troitsk, Moscow 142190, Russia}
\author{Viatcheslav N. Agafonov}
\affiliation{GREMAN, UMR CNRS CEA 7347, University of Tours, 37200 Tours, France}
\author{Wolfram H. P. Pernice}
\affiliation{Institute of Physics and Center for Nanotechnology, University of Münster, D-48149 Münster, Germany}
\author{Alexander Kubanek}
\email{alexander.kubanek@uni-ulm.de}
\affiliation{Institut f\"ur Quantenoptik, Universit\"at Ulm, D-89081 Ulm, Germany}
\affiliation{Center for Integrated Quantum Science and Technology (IQst), Ulm University, Albert-Einstein-Allee 11, D-89081 Ulm, Germany}

\newacro{pl}[PL]{photoluminescence}
\newacro{nd}[ND]{nanodiamond}
\newacroplural{nd}[NDs]{nanodiamond}
\newacro{pcc}[PCC]{photonic crystal cavity}
\newacroplural{pcc}[PCCs]{photonic crystal cavity}
\newacro{afm}[AFM]{atomic force microscope}
\newacro{siv}[SiV$^-$]{negatively charged silicon-vacancy center in diamond}
\newacro{opo}[OPO]{optical parametric oscillator}
\newacro{fwhm}[FWHM]{full width at half maximum}

\begin{abstract}

Spin-based quantum photonics promise to realize distributed quantum computing and quantum networks. The performance depends on efficient entanglement distribution, where the efficiency can be boosted by means of cavity quantum electrodynamics. The central challenge is the development of compact devices with large spin-photon coupling rates and high operation bandwidth. Photonic crystal cavities comprise strong field confinement but put high demands on accurate positioning of an atomic system in the mode field maximum. Color center in diamond, and in particular the negatively-charged Silicon-Vacancy center, emerged as a promising atom-like systems. Large spectral stability and access to long-lived, nuclear spin memories enabled elementary demonstrations of quantum network nodes including memory-enhanced quantum communication. In a hybrid approach, we deterministically place SiV$^-$-containing nanodiamonds inside one hole of a one-dimensional, free-standing, Si$_3$N$_4$-based photonic crystal cavity and coherently couple individual optical transitions to the cavity mode. We optimize the light-matter coupling by utilizing two-mode composition, waveguiding, Purcell-enhancement and cavity resonance tuning. The resulting photon flux is increased by more than a factor of 14 as compared to free-space. The corresponding lifetime shortening to below \SI{460}{\pico\second} puts the potential operation bandwidth beyond GHz rates. Our results mark an important step to realize quantum network nodes based on hybrid quantum photonics with SiV$^-$- center in nanodiamonds. 

\end{abstract}

\maketitle

Quantum technologies offer revolutionary capabilities in the transmission and processing of information with built-in security, based on the principles of quantum mechanics. The realization of a multi-purpose quantum network is a desired goal in the field of quantum communications and computing that would be capable of simultaneously implementing many applications including secure key distribution \cite{lo_2014_SecureQuantumKey,scarani_2009_SecurityPracticalQuantum}, quantum internet \cite{kimble_2008_QuantumInternet,pant_2019_RoutingEntanglementQuantum}, distributed and remote-access quantum computing \cite{meter_2016_PathScalableDistributeda,beals_2013_EfficientDistributedQuantum}, covert communications and blind quantum computation \cite{sheikholeslami_2016_CovertCommunicationClassicalquantum,fitzsimons_2017_PrivateQuantumComputation}, clock synchronization \cite{komar_2014_QuantumNetworkClocks} and longer baseline telescopes \cite{gottesman_2012_LongerBaselineTelescopesUsing}. The key element of a quantum network is the quantum node. Its task is to mediate interaction and to transfer quantum information between the stationary qubits (spins) and flying qubits (photons). One key benchmark of the quantum node is the operational bandwidth. High-bandwidth spin-photon interfaces are realized in the field of cavity quantum electrodynamics (cQED) where atomic transitions are coupled to the mode of an optical resonator. The excited state lifetime of the transition is drastically shortened in the regime of strong Purcell enhancement resulting in a significant efficiency boost of the quantum node. 

Here, we report Purcell enhancement and lifetime shortening by more than an order of magnitude. Therefore, we use a hybrid quantum photonics approach that allows us to integrate atomic systems in complex, on-chip photonic circuits \cite{kim_2020_HybridIntegrationMethods}. Hybridization to date often relies on evanescent coupling \cite{bohm_2019_OnchipIntegrationSingle, wolters_2010_EnhancementZeroPhonon, fehler_2020_PurcellenhancedEmissionIndividuala, englund_2010_DeterministicCouplingSingle} with coupling rates limited by the exponential decay of the electric field. Our approach is based on a long-standing goal in cQED, namely to position an atom in the electric field maximum of a photonic crystal cavity (PCC). However, the approach requires highly accurate position control of the atom within the center of a PCC hole. First concepts were developed for single neutral atoms in combination with hybrid atom trapping including optical and Casimir-Polder forces \cite{hung_2013_TrappedAtomsOnedimensional}. Recent developments have tackled this challenge with nano-pockets \cite{froch_2020_PhotonicNanobeamCavities, alagappan_2018_DiamondNanopocketNew} or trenched nanobeam cavities \cite{alagappan_2020_PurcellEnhancementLight}. In our approach, we position an atom-like quantum emitter, namely a \siv center, which is located inside a nanometer-sized solid-state host, namely a nanodiamond (ND), with high accuracy inside the hole of a one-dimensional, free-standing \ac{pcc} \cite{schrinner_2020_IntegrationDiamondBasedQuantum}. The high refractive index of the diamond host ($n\tief{Dia}=2.4$) is comparable with the refractive index of the PCC material ($n\tief{Si_3N_4}\approx 2.0$), which leads to an efficiency boost due to waveguiding in addition to Purcell-enhancement. 

We begin our work with numerical simulations individually discussing the effect of waveguiding and Purcell enhancement and disclose that waveguiding is only sufficiently strong for dipoles placed inside the waveguide. Experimentally, we begin with AFM-based nanomanipulation in order to deterministically place \cite{schell_2011_ScanningProbebasedPickandplace} a precharacterized single ND with incorporated \siv centers within one hole of the \ac{pcc}. We utilize two-mode composition to enable further improvement in coupling strength as compared to a single mode. By tuning the effective cavity mode in resonance with individual transitions of the \siv  center we show \ac{pl} enhancement of up to 14 and shortened lifetimes of individual transitions of \SI{460}{\pico\second} due to the Purcell-effect.

\section{Simulation}
\begin{figure}
\centering
\includegraphics[scale=.95]{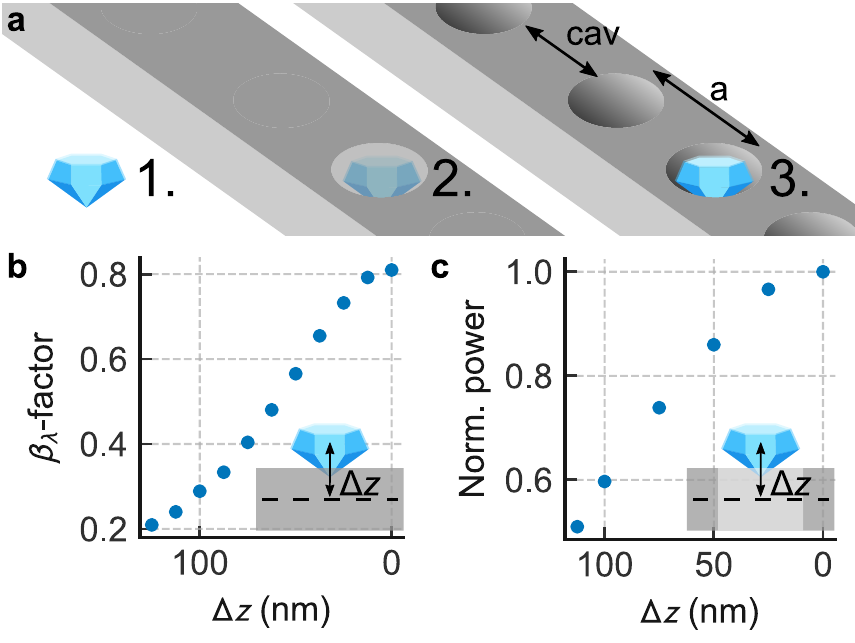}
\caption{Simulation of emitter and waveguide/\ac{pcc} \textbf{a} Three different cases are taken into account. First, the emitter in vacuum. Second, the emitter placed inside the waveguide without holes and third, the emitter placed inside the second hole of the \ac{pcc} at the same location as in the second case. For the simulation, the position of the emitter along the $z$-axis of the waveguide is varied. \textbf{b} $\beta_\lambda$-factor at \SI{737}{\nano\meter} of an emitter coupled to a waveguide versus distance of the emitter to the center of the beam. As the emitter is moved closer to the center of the beam, the $\beta_\lambda$ factor increases from  0.2 to above 0.8 \textbf{c} Simulation of coupled power against the distance from the emitter to the cavity axis. The cavity is simulated to possess a mode at $\approx\SI{740}{\nano\meter}$ and the emitter is shifted from zero displacement to over \SI{100}{\nano\meter} away from the cavity axis. A drop in coupled power of almost \SI{50}{\percent} can be observed.}
\label{fig:pcc_for_simulation}
\end{figure}

The quantum emitter's spatial degrees of freedom are one of the most crucial factors influencing the coupling strength and often hinders high Purcell-factors. Evanescent coupling of color center in NDs to waveguides can offer coupling on the single photon level \cite{bohm_2019_OnchipIntegrationSingle}. Nevertheless, an emitter inside the waveguide can show stronger coupling due to better electric field overlap. We perform FDTD simulations based on the software MEEP \cite{oskooi_2010_MeepFlexibleFreesoftware} in order to show the dependence of the coupling strength on the emitter position. Therefore, we vary the position along the $z$-axis of the waveguide (see Fig.\,\ref{fig:pcc_for_simulation}a, b).

For normalization purposes, we first simulate the flux of a dipole source in vacuum with a gaussian spectral profile through a surrounding box. Subsequently, we calculate the emission of the same dipole source placed inside a waveguide. In Fig.\,\ref{fig:pcc_for_simulation}a the first and second case show a schematic of respective arrangements. Considering the photon flux through two planes at the end of the waveguide distinguishes the coupled photons from the uncoupled ones. This flux is divided through the total emission of the emitter. We introduce the spectrally resolved $\beta_\lambda$ factor \[\beta_\lambda = \frac{\Gamma\tief{wave}(\lambda)}{\Gamma\tief{wave}(\lambda)+\Gamma\tief{0}(\lambda)},\]
to quantify the coupled emission at a certain wavelength, where $\Gamma\tief{wave}(\lambda)$ is the emission rate into the waveguide at a certain wavelength $\lambda$ and $\Gamma\tief{0}(\lambda)$ is the uncoupled emission rate at the same wavelength. In Fig.\,\ref{fig:pcc_for_simulation}b the $\beta_\lambda$-factor at \SI{737}{\nano\meter} is shown \textit{versus} the emitter's $z$-position. The highest coupling to the waveguide is achieved for placing the emitter at the center position of the waveguide. A simulated $\beta_\lambda$-factor of $\approx 0.8$ can be achieved for a perfectly aligned dipole. If the coupled emission should be further increased, two approaches can be pursued: Either decreasing the uncoupled emission by reducing the probability of the uncoupled channel $\Gamma_0$
\cite{arcari_2014_NearUnityCouplingEfficiency}
or increasing the coupled emission (as in our case by introducing a cavity). In the last simulation step, the emitter sits at the same position as in the previous case, but inside an air hole of the cavity. The cavity consists out of two bragg mirrors, formed by alternating dielectric medias along the waveguide. In Fig.\,\ref{fig:pcc_for_simulation}c the power of an emitter, coupled to a cavity mode at approximately $\SI{740}{\nano\meter}$ is plotted against the displacement of the emitter to the cavity axis. The emitted power decreases by a factor of two when the emitter is displaced by about $\SI{100}{\nano\meter}$.
Summarizing, the emitter needs to be placed inside the hole of the \ac{pcc} in order to achieve maximal emitter-cavity coupling. Therefore, we need to pick a \nd with the same size as the diameter of the \ac{pcc}'s hole and deterministically position it within the hole.
\section{Results}
\subsection{Device and post-processing}
Our photonic interface is based on a freestanding one-dimensional \ac{pcc} built out of Si$_3$N$_4$
\cite{fehler_2019_EfficientCouplingEnsemble, fehler_2020_PurcellenhancedEmissionIndividuala}.
 The cavity is formed by two bragg mirrors with N=43 holes separated from each other with a period of $a = \SI{265}{\nano\meter}$, while the cavity defect is introduced between the two bragg mirrors with a distance of \SI{232}{\nano\meter}. The waveguide beam is freestanding and hence yields a high refractive index contrast of the cavity holes and additionally minimizes coupling of emission to the base substrate. This probe waveguide is optimized for \SI{740}{\nano\meter} to match the ZPL of the \siv. The cavity transmission is directed out of plane by grating couplers connected to the probe beam enabling detection of fluorescence with a confocal microscope.
 
\begin{figure*}
\centering
\includegraphics[scale=.95]{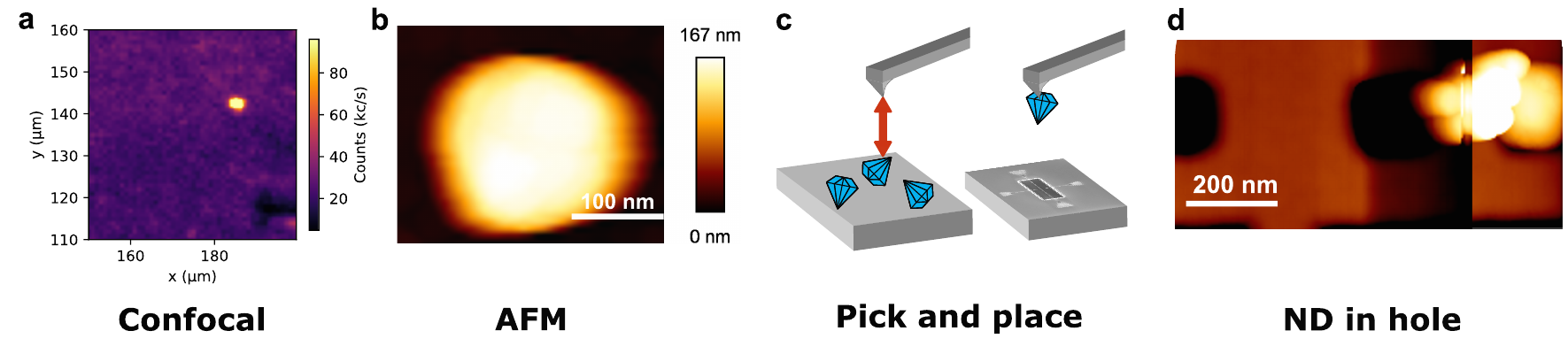}
\caption{Pick and place procedure. \textbf{a}~In a confocal microscope setup, \acp{nd} with \acp{siv} are investigated and spectrally precharacterized. \textbf{b}~After the optical characterization, the substrate is scanned in an \ac{afm} and manually overlapped with the confocal image to find the desired \nd. \textbf{c}~To pick the emitter from the substrate, the cantilever approaches the \nd multiple times with a constant force. Afterwards, the sample is changed to the photonic chip and the region of interest is carefully scanned with the \nd attached to the cantilever. \textbf{d}~\ac{afm} scan of the cavity region after the \nd was placed in the second hole. Between the first and second hole, residues from the coating of the cantilever or of the \nd host dispersion were also stripped off.}
\label{fig:pick_and_place}
\end{figure*}
The first step in the functionalization of the \ac{pcc} is to identify \acp{nd} of appropriate size containing \siv center with the right spectral properties. For the precharacterization, we coat a dispersion of \siv -containing \acp{nd} onto a substrate and then investigated individual sites in a home-built confocal microscope at cryogenic temperatures. Fig.\,\ref{fig:pick_and_place}a shows a confocal scan with an \ac{nd} containing \siv centers. To reach highest coupling for individual lines, we choose an ensemble of \siv center with spectrally distributed transition frequencies due to internal strain
\cite{rogers_2019_SingleSiVCenters}. In the next step, we follow a modified pick and place procedure, where the substrate is scanned with the help of an \ac{afm} and overlapped with the low temperature confocal image to ensure the correct emitter is found (Fig.\,\ref{fig:pick_and_place}b). The \nd is picked up by applying a constant force with the platin-coated cantilever tip \cite{schell_2011_ScanningProbebasedPickandplace}. With the attached \nd and the photonic chip mounted in the \ac{afm} (Fig.\,\ref{fig:pick_and_place}c), the cavity can be located by non-contact scanning. Afterwards, we place the \nd in the desired region within the \ac{pcc}. In Fig.\,\ref{fig:pick_and_place}d the placed \nd is located in the second hole of one of the bragg mirrors. The particles between the holes are residues from \nd host dispersion or cantilever tip coating.

\subsection{Cavity Emitter Coupling}
The spontaneous emission (SE) rate enhancement can be described with the help of the Purcell-factor \cite{englund_2005_ControllingSpontaneousEmission}:
\begin{equation}
F\tief{cav} = 
\frac{3}{4\pi^2}\frac{\lambda^3}{n^3}\frac{Q}{V},
\end{equation}
where $\lambda$ is the central wavelength, $n$ is the refractive index, $Q$ is the quality factor, $V$ is the mode volume. If the emitter is spatially and spectrally detuned from the cavity modes, further terms need to be introduced to obtain the overall enhancement of the emission:
\begin{equation}
\frac{\Gamma\tief{cav}}{\Gamma_0} = \sum_{k} F\tief{\textit{k}, cav}
\underbrace{
\left( \frac{ \vec{E_k} (\vec{r})  \cdot \vec{\mu} }{|\vec{E_k}\tief{, max}| |\vec{\mu}|} \right)^2}_\mathrm{positioning}
\underbrace{\frac{1}{1+4Q_k^2(\frac{\lambda\tief{SiV}}{\lambda_k}-1)^2}}_\mathrm{spectral\, mismatch}
\label{eq:se_enhancement}
\end{equation}
where $\Gamma\tief{cav}$ is the emission rate into the cavity, $\Gamma_0$ is the free-space emission of the emitter, $\vec{E}_k(\vec{r})$ is the electric field at position $\vec{r}$ while $k$ accounts for the k$^\mathrm{th}$ cavity mode, $\vec{\mu}$ is the dipole moment of the emitter and $\lambda\tief{SiV}$ is the wavelength of the \siv transition. There are three crucial terms for the enhancement: The Purcell-factor $F\tief{cav}$, the spatial positioning of the emitter in the field of the cavity mode as well as the spectral mismatch of the cavity mode with respect to the emitter's transition frequency. 

The first term, $F\tief{cav}$, can be controlled by optimizing the \ac{pcc} in simulations prior to fabrication. For example, by optimizing the geometry to achieve the highest $Q$-factors and small mode volumes $V$. To achieve high quality \acp{pcc}, the waveguide of the \ac{pcc} is free-standing, such that the loss of electric field in the base substrate is at minimum and to have a high refractive index contrast of the cavity holes.

The second term of Eq.\,\ref{eq:se_enhancement} describes the spatial positioning of the emitter with respect to the electric field of the cavity mode as well as the alignment of the dipole to the cavity polarization which can be controlled by \ac{afm} nanomanipulation \cite{haussler_2019_PreparingSingleSiV}.

\begin{figure*}
\centering
\includegraphics[width=\textwidth]{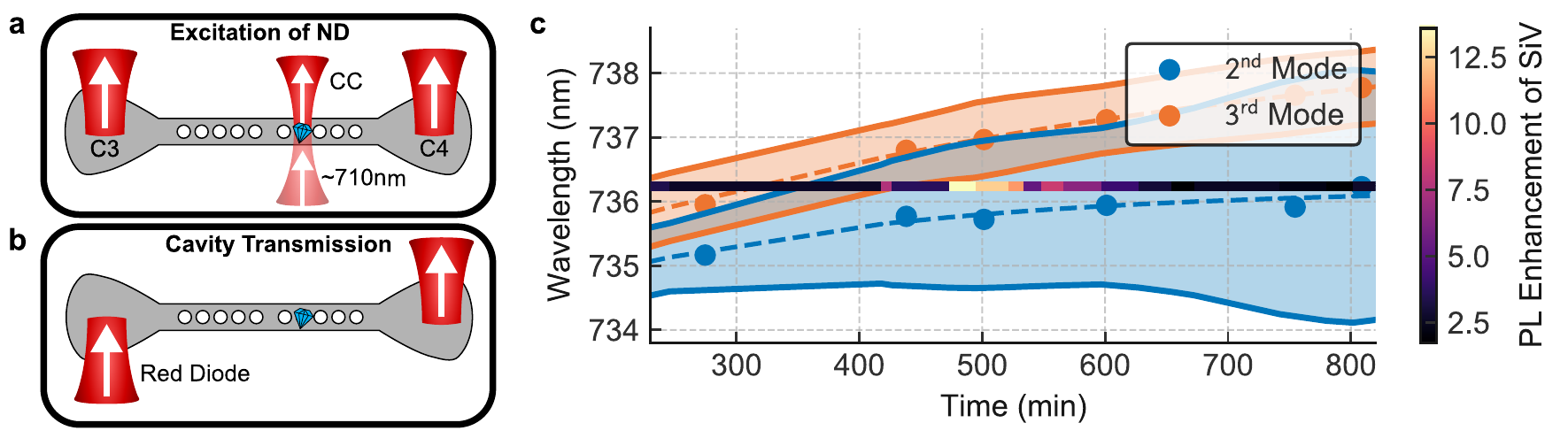}
\caption{Readout of coupled emitter. \textbf{a}~Off-resonant excitation of emitter \textit{via} scattering in an cavity mode at $\approx\SI{710}{\nano\meter}$. The emitter is directly excited at its center position and can be read out at port 3 (C3), its center position (CC) and at port 4 (C4). \textbf{b}~Cavity transmission measurement, \textit{via} a spectrally broad red diode, coupled into port 3 and detected at port 4 (34). \textbf{c}~The \second (blue) and \third (orange) cavity mode shows an exponential shift over time due to gas freezing to the \ac{pcc}. The dashed line indicates a fit through the center position of each mode, while the solid line corresponds to the FWHM. The \siv transition at \SI{736.2}{\nano\meter} couples to both modes, resulting in a higher coupling between the modes. The photoluminescence of the \siv transition is enhanced by a factor of 14.}
\label{fig:measurement_methods_and_pl_enhancement}
\end{figure*}

The third term describes the spectral mismatch between the cavity mode resonance and the emitters transition frequency. While there is ongoing research to realize intrinsically matched emitter-cavity systems \cite{ondic_2020_PhotonicCrystalCavityenhanced} most systems rely on spectral matching  controlled, for example, by freezing gas to the \ac{pcc} at cryogenic temperatures \cite{mosor_2005_ScanningPhotonicCrystal}. For higher order cavity modes the electric field is further extended into the bragg mirror, while for the first mode the maximum is located in the center of the cavity. As it can be seen in the Supplementary Information, the insertion of a \nd inside a hole drastically alters the field distribution of the cavity modes due to a changed refractive index. In consequence, we place the \nd inside the second hole of the \ac{pcc} enhancing the coupling to higher-order modes which have lower $Q$-factors and larger mode extend. However, they are less prone to spectral mismatch. 

The spectral mode position of the cavity can be probed by observing the cavity transmission signal of a broadband diode ($\lambda\tief{central}\approx \SI{730}{\nano\meter}$). Therefore, we couple light into port 3 and measure the transmission at port 4 as shown in Fig.\,\ref{fig:measurement_methods_and_pl_enhancement}b. In Fig.\,\ref{fig:measurement_methods_and_pl_enhancement}c, the spectral positions of the \second and \third modes are plotted as a function of time while the \ac{pcc} is cooled in a continuous flow-cryostat (Janis ST-500) to temperatures of approximately $\SI{2.5}{\kelvin}$. Residual gas inside the cryostat is continuously freezing to the \ac{pcc} and changes the effective refractive index. This leads to a shift in the central wavelength of each mode to higher wavelengths. The \second mode shows a slightly slower drift with respect to the \third mode, while the freezing process has a larger influence on the full width at half maximum (FWHM) of the \second mode. Both modes can be shifted by more than $\SI{2}{\nano\meter}$, which enables controlled and time resolved coupling to individual transitions. Stabilization of the modes resonance frequency could be achieved via local heating of the \ac{pcc}.

The excitation for \ac{pl} measurements is performed by off-resonant pumping at $\SI{710}{\nano\meter}$ with a continuous wave titan-sapphire laser at the position of the emitter. The laser scatters into a cavity mode at shorter wavelength in comparison to the emitter transition frequency yielding efficient excitation of the \siv center. In order to ensure high excitation efficiency, we spectrally overlap the local mode of the \siv with the blue detuned cavity mode  \cite{rogers_2014_ElectronicStructureNegativelya} (see Supplementary Information for further details about excitation scheme). Therefore, we operate at low excitation powers of $\approx$\SI{50}{\micro\watt}, which ensures low background fluorescence of the Si$_3$N$_4$ host material. We collect cavity coupled emission from the \siv center at port 3 and 4. Fig.\,\ref{fig:measurement_methods_and_pl_enhancement}a displays a schematic of the measurement procedure. As the emitter is excited at the center position and read out at port 3 and 4, these two emission channels are labeled as C3 and C4. We determine the freespace emission by exciting and collecting the intensity at the emitter position (CC) weighted by setup efficiency. We then extract the PL enhancement \textit{via} \[\frac{\Gamma\tief{cav}}{\Gamma\tief{free}} = \frac{\Gamma\tief{C3} + \Gamma\tief{C4}}{\Gamma\tief{CC}} \frac{\eta\tief{free}}{\eta\tief{cav}},\]
where $\Gamma\tief{free}=\Gamma\tief{CC}/\eta\tief{free}$ is the emission rate into freespace and $\Gamma\tief{C3, C4}$ are the rates from  C3, C4 respectively. The factors $\eta\tief{free, cav}$ correspond to the detection efficiency of the cavity channel as well as the freespace channel respectively. 
We now make use of composite cavity modes in order to further increase the emitter-cavity coupling strength. Since the emitter couples to more than one mode which are very closely spaced and hence overlap, the most effective coupling is observed between the two overlapping modes considered herein (see Supporting Information). We can see from Eq.\,\ref{eq:se_enhancement}, that the PL enhancement is proportional to the $Q$-factor and its detuning from the transition line
\begin{equation}
\frac{\Gamma\tief{cav}}{\Gamma\tief{free}} \propto \sum_{k} \frac{Q_k}{1+4Q_k^2(\lambda\tief{SiV}/\lambda_k-1)^2} = \sum_{k} Q_{k, \mathrm{w}}.
\end{equation}
Here, $Q_{k, \mathrm{w}}$ is defined as the weighted $Q$-factor of the k$^\mathrm{th}$-mode according to the spectral detuning from the emitter. However, for a more complete analysis, one should also consider the difference of the mode volume and the product of electric field with dipole moment $\vec{E_k} (\vec{r})  \cdot \vec{\mu}$. As we tune the cavity over the transition of the \siv, its PL intensity is enhanced up to a factor of 14. 

\begin{figure*}
\centering
\includegraphics[scale=.9]{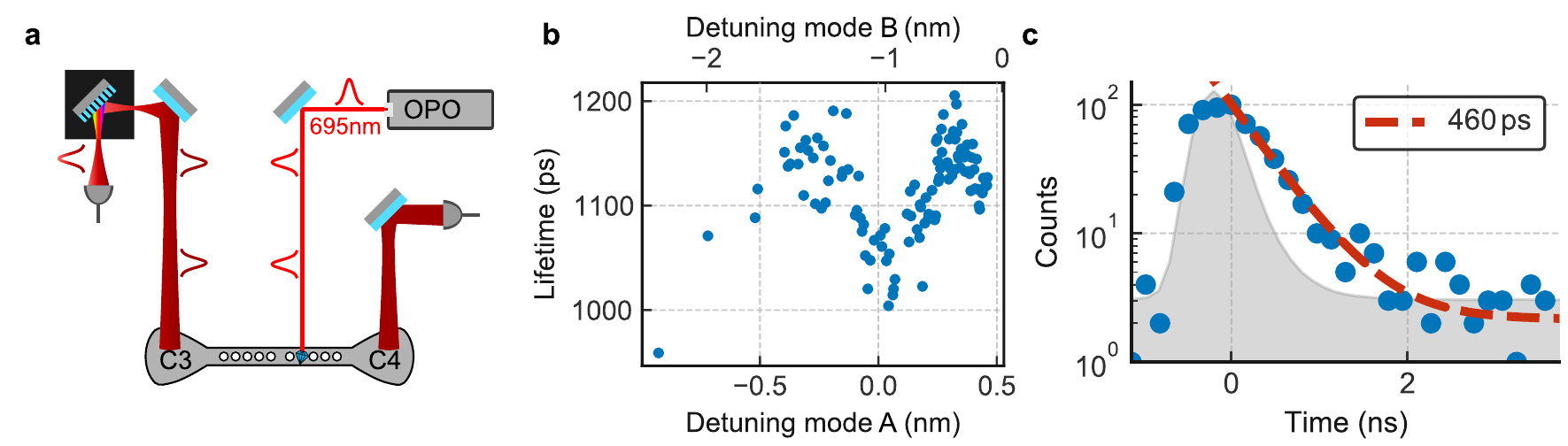}
\caption{Lifetime measurement: \textbf{a}~Setup for pulsed lifetime measurements of single \siv lines. The \nd is enhanced by a pulsed \opo from top. The coupled emission is collected from one port (C3) and spectrally filtered through a commercial spectrometer. Only a single line of the \siv passes the output slit and is directed on a detector. The other output port (C4) is used for optimization during the measurement. \textbf{b}~Tuning of the split-up \first mode (mode A and B) of the \ac{pcc} over the \siv transition at $\approx\SI{736}{\nano\meter}$. A clear reduction in lifetime is visible as a function of cavity detuning. The latter is deduced from exponential exponential fits to transmission measurements, as sketched in Fig.\,\ref{fig:measurement_methods_and_pl_enhancement}b. \textbf{c} Lifetime measurement of a line at $\approx\SI{733}{\nano\meter}$ with a lifetime of $\SI{460}{\ps}$. The grey shadowed area corresponds to the smallest time detectable with our equipment.}
\label{fig:lifetime}
\end{figure*}

The emitter-cavity coupling also manifests in a shortened lifetime due to the Purcell effect
 \cite{zhang_2018_StronglyCavityEnhancedSpontaneous}:
\begin{equation}
F\tief{P} = \left(\frac{\tau\tief{off}}{\tau\tief{on}}-1\right)/\xi,
\end{equation}
where $\tau\tief{off, on}$ are the off (on) resonant lifetimes and $\xi$ is the off-resonant branching ratio (see Supplementary Information).
We directly excite the \siv center with a pulsed \acf{opo} and record the fluorescence decay through the cavity channel. Fig.\,\ref{fig:lifetime}a shows a sketch of the measurement setup. The excitation wavelength at \SI{695}{\nano\meter} with a \ac{fwhm} of $\mathrm{FWHM} \approx\SI{9}{\nano\meter}$ is far-detuned from the \siv 's ZPL but resonant with a blue-detuned mode of the \ac{pcc} as described above. One output port (C4) of the cavity is used for real-time position optimization of pump light location while the emission out of the other port (C3) is narrow-band filtered ($\ac{fwhm} \approx\SI{60}{\giga\hertz}$) through a commercial spectrometer (Princeton Instruments HRS-500). Only one specific transition frequency passes the spectrometer slit and is directed on a single photon detector connected to a acquisition device (Swabian Instruments Time Tagger Ultra). 
Again, gas tuning allows us to resolve detuning-dependent coupling revealed by changing lifetime measurements of the same transition discussed in Fig.\,\ref{fig:measurement_methods_and_pl_enhancement}c. 

From cavity transmission measurements, carried out in between subsequent lifetime measurements, we extrapolate the lifetime versus cavity mode detuning (Fig.\,\ref{fig:lifetime}b). Our experimental time-resolution is limited by the timing-jitter of the single photon counting module (Excelitas AQRH-14-FC) to about \SI{250}{\pico\second}. The lifetime of the highest coupling in Fig.\,\ref{fig:measurement_methods_and_pl_enhancement}c corresponds to a lifetime approximately within this temporal limit. 
Transmission measurements of the \first mode reveal that this mode is also composed out of two sub-modes. For simplicity, we will refer to them as mode A and B. 
It is apparent in Fig.\,\ref{fig:lifetime}b that coupling is higher for larger negative detunings where the \second and \third  cavity modes are located. 

The shortest lifetime measured on a different filtered \ac{siv} transition was as low as \SI{460}{\pico\second}. The corresponding PL evaluation yields a PL-enhancement of $4$ in agreement with the reduced lifetime. Fig.\,\ref{fig:lifetime}c shows the lifetime measurement together with the system response when a bare ps laser pulse is applied (grey area) indicating the resolution limit of our setup. The off-resonant lifetime $\tau\tief{off}$ can not be reliably deduced in neither of the two cases (Fig.\,\ref{fig:lifetime}b and c) due to non-vanishing mode overlap and the change of the dielectric environment. A detailed discussion of this subject is given in the Supplementary Information. 

\section{Outlook}
Summarizing, by positioning a \ac{siv} containing ND inside the hole of a PCC and by taking advantage of two-mode composition, cavity resonance tuning, Purcell-enhancement and waveguiding we reach enhanced photon rates by more than a factor of 14 as compared to free space emission. The resulting lifetime shortening below \SI{460}{\pico\second} puts 
our platform among the few spin-based photonics systems with bandwidth beyond GHz-rates. Shorter lifetimes are beyond the time resolution of our detection setup. We use the measured lifetime as benchmark to compare our hybrid system with the best performing competing platform, in particular \siv center coupled to the mode of an all-diamond photonic crystal cavity. Reported lifetimes between \SI{150}{\pico\second} and \SI{194}{\pico\second} \cite{zhang_2018_StronglyCavityEnhancedSpontaneous} are comparable to the lifetime we expect for our measured PL enhancement of 14 (although beyond our detection resolution). The resulting Purcell factor relies on the branching ratio and on the off-resonant lifetime. In reference \cite{zhang_2018_StronglyCavityEnhancedSpontaneous} a Purcell enhancement of 26.1 results from a branching ratio of transition B of 32.5 \% and from an off-resonant lifetime of \SI{1.84}{\nano\second}. Our measured lifetime of \SI{460}{\pico\second} together with a minimum off-resonant lifetime of \SI{1.2}{\nano\second} results in an off-resonant branching ratio of 0.4 which agrees well with the range between 0.031 and 0.452 as reported in \cite{zhang_2018_StronglyCavityEnhancedSpontaneous}. Details on the evaluation are given in the Supplementary Information. Our hybrid architecture brings the advantages of a well-established photonics platform and is compatible with large scale integration into photonic integrated circuits \cite{wan_2020_LargescaleIntegrationArtificial}. Considering the \siv center's exceptional spectral stability and narrow inhomogeneous distribution \cite{becker_2017_CoherencePropertiesQuantum, rogers_2019_SingleSiVCenters} as well as the capability to access  long-living, nuclear-spin quantum memories in the diamond host \cite{metsch_2019_InitializationReadoutNuclear} makes our system a very promising candidate for spin-based quantum technologies 
\cite{atature_2018_MaterialPlatformsSpinbased, awschalom_2018_QuantumTechnologiesOptically} and in particular efficient quantum nodes in large-scale quantum networks \cite{bhaskar_2020_ExperimentalDemonstrationMemoryenhanced}. 

\section{Methods}
\subsection{Nanodiamond Fabrication}
\begin{figure}[htpb]
\centering
\includegraphics[width=.5\columnwidth]{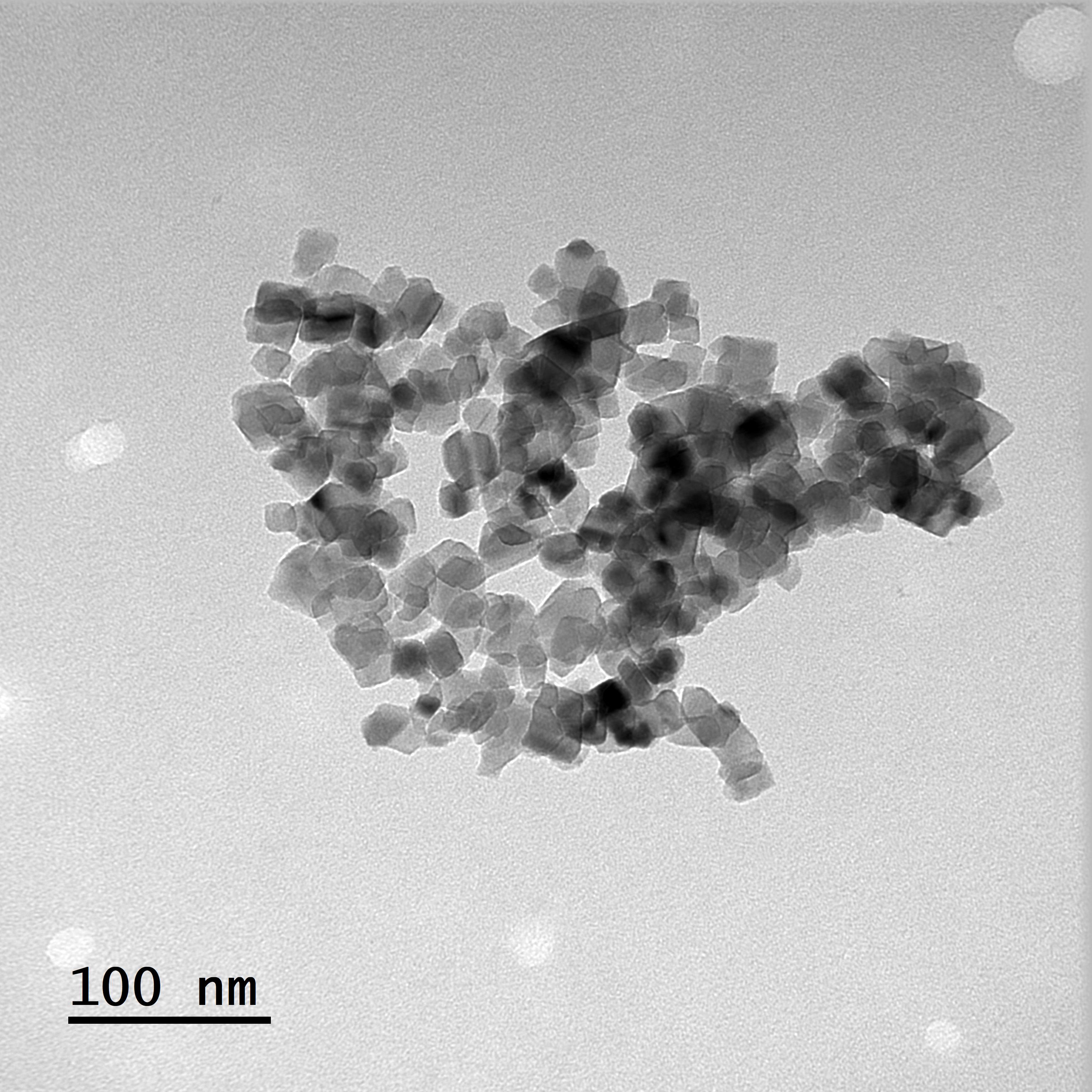}
\caption{TEM image of \nd cluster with incorporated \siv centers.}
\label{fig:nd_tem}
\end{figure}

Nanodiamonds with \siv centers were produced by high pressure – high temperature (HPHT) synthesis based on the hydrocarbon metal catalyst-free growth system presenting homogeneous mixtures of naphthalene C$_{10}$H$_8$ (Chemapol),  detonation ultrananosized (3-\SI{5}{\nano\meter}) diamonds (SkySpring Nanomaterials) and tetrakis(trimethylsilyl)silane (C$_{12}$H$_{36}$Si$_5$) with the initial atomic silicon-to-carbon (Si/C) ratio = 1/100. The synthesis was performed on a high-pressure apparatus of "Toroid" type \cite{davydov_2014_ProductionNanoMicrodiamonds}. The experimental procedure consists of loading the high-pressure apparatus to \SI{8.0}{\giga\pascal}, heating the samples up to \SI{1450}{\celsius} and short (\SI{3}{\second}) isothermal exposures at these temperatures. The chemical purification of diamond materials was carried out by processing them with a mixture of acids (HNO$_3$-HClO$_4$-H$_2$SO$_4$). Aqueous dispersion of nanodiamonds were obtained with the help of ultrasonic UP200Ht dispersant (Hielscher Ultrasonic Technology). The synthesized nanodiamonds were examined by the transmission electron microscopy (TEM). A typical TEM image of the received nanodiamonds is presented on Fig. \ref{fig:nd_tem}.

\subsection{Fabrication of Photonic Devices}

Experimental realization of free-standing PCC devices is performed on Silicon nitride-on-insulator wafers comprising of \SI{200}{\nano\meter} stoichiometric Si$_3$N$_4$ on top of a \SI{2}{\micro\meter} thick SiO$_2$ layer on top of Si. The fabrication of the nanophotonic circuits is carried out using several steps of electron-beam lithography. In the first step, the nanophotonic devices are defined on top of the Si$_3$N$_4$ layer by exposing negative photoresist ma-N 2403. Subsequently, the devices were \SI{75}{\percent} dry-etched into the silicon nitride layer using a CHF$_3$/O$_2$ plasma. Freestanding PhC cavity design is obtained by removing underneath SiO$_2$. Thus, the second step includes opening the window around the PhC region by means of exposing positive photoresist PMMA in this area. In the next step, the remaining \SI{25}{\percent} of silicon nitride in the window area is etched, while the waveguide inside of the window is protected with an ma-N 2403 photoresist, whereas the waveguides outside the windows are protected by unexposed PMMA mask. After this step, both photoresists were removed by O$_2$ plasma. Suspending of the cavity region was obtained by removing  SiO$_2$ layer in the windows by wet etching, namely by immersing the chip in hydrofluoric acid (HF).

\subsection{Optical Methods}
Our optical setup consists out of a home-build confocal microscope with a 0.55 NA objective, running on the software Qudi \cite{binder_2017_QudiModularPython}. The excitation is performed using a Titan:Sapphire Laser for the continuous measurements and a pulsed \ac{opo} for all pulsed measurements. Two additional paths within the confocal setup allow a fixed excitation position on the sample with simultaneous detection of the cavity outputs from both couplers. The lifetime measurements are counted with an avalanche photo diode (Excelitas AQRH-14-FC) and directed either on the time tagger QuTau from qutools or on the Time Tagger Ultra from Swabian Instruments. For the spectral filtered measurements the emission is directed through the spectrometer HRS-500 from Princeton Instruments.

\section{Acknowledgments}
The project was funded by the Baden-Württemberg Stiftung in project Internationale Spitzenforschung. AK acknowledges support of the BMBF/VDI in project Q.Link.X, the European fund for regional development (EFRE) program Baden-Württemberg and the Deutsche Forschungsgemeinschaft (DFG, German Research Foundation) in project 398628099. KGF and AK acknowledge support of IQst. The AFM was funded by the DFG. We thank Prof. Kay Gottschalk and Frederike Erb for their support. VAD thanks the Russian Foundation for Basic Research (grant No. 18-03-00936) for financial support.

\providecommand{\latin}[1]{#1}
\makeatletter
\providecommand{\doi}
  {\begingroup\let\do\@makeother\dospecials
  \catcode`\{=1 \catcode`\}=2 \doi@aux}
\providecommand{\doi@aux}[1]{\endgroup\texttt{#1}}
\makeatother
\providecommand*\mcitethebibliography{\thebibliography}
\csname @ifundefined\endcsname{endmcitethebibliography}
  {\let\endmcitethebibliography\endthebibliography}{}

\end{document}